\documentclass[preprint2]{aastex}

\shorttitle{DIG in the Milky Way}
\shortauthors{Elwert & Dettmar}

\begin{document}

\title{Constraining the extra heating of the Diffuse Ionized Gas in the Milky Way}

\author{T. Elwert\altaffilmark{1} and R.--J. Dettmar}
\affil{Astronomisches Institut, Ruhr--Universit\"at Bochum,
             Universit\"atsstr. 150, D--44780 Bochum, Germany}
\email{elwert@pa.uky.edu, dettmar@astro.rub.de}
\altaffiltext{1}{present address: Department of Physics and Astronomy, 
        University of Kentucky, Lexington, KY 40506, USA}

\begin{abstract}

% change Feb. 05
The detailed observations of the diffuse ionized gas through the emission lines H$\alpha$, [NII], and [SII] in the Perseus Arm of our Galaxy 
by the Wisconsin H$\alpha$ Mapper (WHAM)--survey challenge photoionization models. They have to explain the observed rise in
the line ratios [NII]/H$\alpha$ and [SII]/H$\alpha$. 
The models described here consider for the first time the detailed observational geometry 
toward the Perseus Arm. 
The models address the vertical variation of the line ratios up to height of 2 kpc above the midplane.
The rising trends of the line ratios are matched. The increase in the line ratios is reflected in a rise of the temperature of the gas layer.
This is due to the progressive hardening of the radiation going through the gas.
However an extra heating above photoionization is needed to explain the absolute values. 
Two different extra heating rates are investigated which are proportional to $n^0$ and $n^1$.
The models show that a combination of both are best to explain the data, where the extra heating 
independent of density is dominant for z $>$ 0.8 kpc. 

\end{abstract}

\keywords{radiative transfer, methods: numerical, Galaxy: halo, ISM: structure}

\section{Introduction}
Containing typically half the mass of ionized hydrogen in galaxies, the
Diffuse Ionized Gas (DIG)  is visible as an extended 
H$\alpha$ emitting layer in our Galaxy (e.g. the WHAM--survey) and in
many other galaxies (see e.g.
T\"ullmann \& Dettmar \cite{tuellmannB},
Collins \& Rand \cite{collins}, Otte et al. \cite{otte},
Hoopes \& Walterbos \cite{hoopes}).
Studies of emission line ratios such as [NII]/H$\alpha$ and [SII]/H$\alpha$ 
provide
information about the physical conditions of the gas. Simple energy
estimations (Reynolds \cite{reynolds90}, Reynolds \cite{reynolds93}) 
favor O and early B stars to be responsible for most of the DIG. 3D models using various methods (e.g. Miller \& Cox \cite{miller2},
Dove \& Shull \cite{dove}, Wood \& Loeb \cite{wood2}, Ciardi et al. \cite{ciardi}, 
Wood et al. \cite{wood}) showed  
that it is possible for ionizing photons from O stars to penetrate from the midplane into the halo.
Wood \& Mathis~\cite{wood3} noted that the line ratios increase with distance from the midplane due
to the progressive hardening of the radiation.
So far photoionization models made no specific attempt to model the trends of the line ratios.
Models by Mathis \cite{mathis}, Domg\"orgen \& Mathis \cite{domgoergen},
Sembach et al. \cite{sembach}, Bland--Hawthorn et al. \cite{bland-hawthorn} used volume average models to explain
the observed data.
The analytical approach by Haffner et al. \cite{haffner}, referred
to as Haffner99 in the rest of this paper, treated the dependence of the line ratios with height.
Haffner99 and its further
application to other galaxies by e.g. Collins \& Rand \cite{collins}, Otte et al. \cite{otte}, 
and Miller \& Veilleux \cite{miller} gave evidence that an additional heating source is needed in 
order to explain the rise of the line ratios with increasing distance $z$ from the midplane. 

We are constructing photoionization models in order to examine the trends in the observed line
ratios and if photoionization can heat up the gas sufficiently in order to explain the data. 
We introduce specific extra heating terms, extra means in addition to photoionization, and 
discuss their properties.
In the following section \ref{data_perseus} we
introduce the observations of the Perseus Arm to which our models
are compared. Section \ref{parameters} 
deals with the model parameters and discusses geometry and
sight line effects taken into account. 
The models are compared with the data in the next section.
The last section summarizes the results.

\section{Data of the Perseus Arm}\label{data_perseus}
We are using the data taken from the Wisconsin H$\alpha$ Mapper (WHAM) 
(e.g.  Haffner99, Haffner et al. \cite{haffner01})  survey which
were kindly provided by Ron Reynolds and Matt Haffner. The WHAM--survey 
mapped the northern sky in H$\alpha$ with declinations of $\delta > -
30^\circ$. The Perseus Arm ($-35^\circ < \delta < -11^\circ$ and $120^\circ < l < 150^\circ$) was additionally mapped in 
[NII]$\lambda$6583 and [SII]$\lambda$6716.
At each pointing an averaged spectrum with a beam of $1^\circ$ is measured
with a velocity resolution of 12 km $\rm s^{-1}$. 
The emission of the Perseus Arm can
be separated in velocity space from the local emission (Haffner99)
for galactic longitudes 
$120^\circ \, < \, \rm l \, < \, 150^\circ$. This was performed by integrating
the line emission in the
velocity range $-100$ km ${\rm s^{-1}}\,<\,v\,< -20$ km $\rm s^{-1}$, 
no line fitting was performed. We are using the intensity of the H$\alpha$ line as well as the 
line ratios [NII]/H$\alpha$ and [SII]/H$\alpha$.
The sensitivity limit of 0.1 Rayleighs\footnote{1 R = $10^6/4\pi$ photons 
$\rm cm^{-2}$ $\rm s^{-1}$ $\rm sr^{-1}$} results in an observed vertical 
height of
up to $|z|$ = 2 kpc assuming a distance to the arm of 2.5 kpc. 

\section{Model parameters}\label{parameters}

We use the spectral simulation code CLOUDY, version 96.00 (described by Ferland  \cite{ferland3}, \cite{ferland1}, 
\cite{ferland2}) to model the DIG. CLOUDY determines the physical conditions by balancing the heating and cooling
rates, so that the energy is conserved. The results of the models are compared to the observed emission line
ratios [NII]/H$\alpha$ and [SII]/H$\alpha$, and the gas temperature as derived from
[NII]/H$\alpha$.

\begin{figure}
   \centering
   \includegraphics[width=5.5cm]{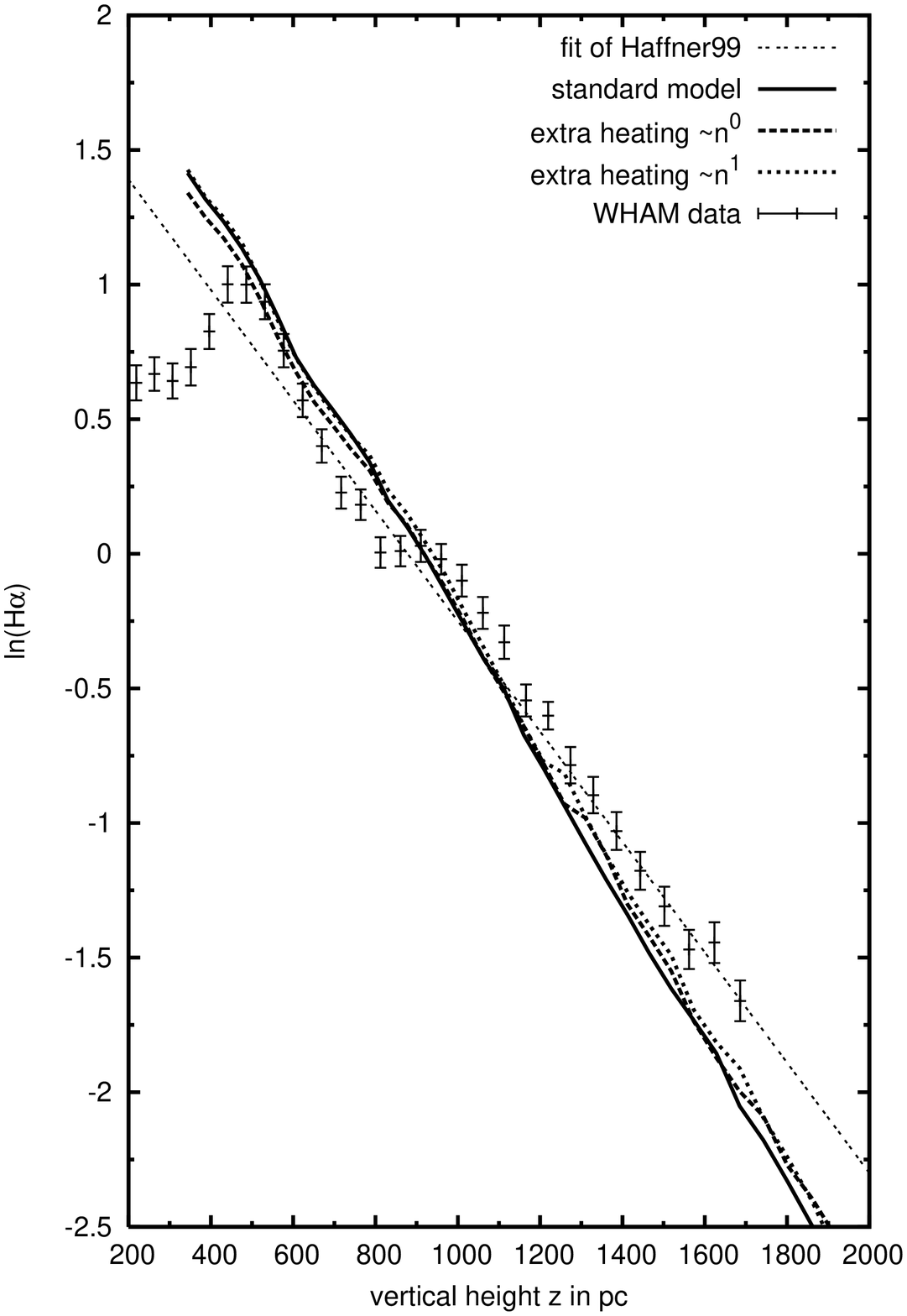}
      \caption{H$\alpha$ intensity of the Perseus Arm as observed by WHAM, overplotted is the
        fit of Haffner99 and the photoionization models with and without extra heating.
              }
         \label{Ha-PerseusArm}
   \end{figure}

In order to realize a model describing the DIG certain parameters have
to be specified: The ionizing spectrum of the source
is a composition of three different stellar temperature: 56\% from 
T = 35000 K, 12\% from T = 40000 K, and 32\% from T = 45000 K, 
as used in Mathis \cite{mathis} and Wood et al. \cite{wood}. 
The WMbasic models (Pauldrach et al. \cite{pauldrach}), which include N--LTE effects, X--ray emission
from shocks within stellar winds, are used as the ionizing spectra. 
The luminosity of the source is chosen in such a way that the observed run of the H$\alpha$ intensity is
matched, as shown in Figure~\ref{Ha-PerseusArm}. 

The density structure is exponential,
as derived from the observed H$\alpha$ intensity, with a scale height of 1 kpc and
a midplane density of 0.2 $\rm cm^{-3}$. The density is in clumps with a filling factor of
20\%, i.e. only 1/5 of the volume is filled with this plasma.
The geometry of the ionized gas is chosen to be a plane parallel layer, i.e. the 
ratio of the depth of the cloud to the distance of the illuminated face to the ionizing 
source is smaller than 1/10. 

We are matching the observed H$\alpha$ intensity of the observations
(Figure~\ref{Ha-PerseusArm}) by placing the illuminated face of the cloud at a z--height of 1 scale
height of the Lockman--layer (300pc) assuming a density law of $n = 0.1 \exp(-z/0.3)$ as in Miller \& Cox~\cite{miller2}.
This is done after the model is calculated as otherwise the condition of plane parallel illumination of the cloud cannot be
fulfilled. The information of the actual position of the ionizing stars are effectively removed, consistent with the
picture of having the DIG being ionized by radiation leaking out of the Lockman--layer. 
The intensity gradient of H$\alpha$ is matched as well as the estimate of the hydrogen ionizing photon
flux by Reynolds~\cite{reynolds90}): $\phi_{DIG} \ge 5 \times 10^6$ hydrogen ionizing photons $\rm cm^{-2}\,\,s^{-1}$.
In order to compare the models on a common basis, which is important as extra heating is dominating 
for large z--heights, we choose to let the ionization structure be the same for all models (see Figure~\ref{ionfrac}). 
This is in accordance with the idea that the extra heating affects only the temperature of the gas. The 
forbidden levels of nitrogen and sulphur can then be more easily excited by collisions with the electrons which
in turn elevates the line ratios [NII]/H$\alpha$ and [SII]/H$\alpha$. 
Figure~\ref{Ha-PerseusArm} shows that the intensity gradient of H$\alpha$ varies only a little for the different
models. The scale heights for the models are slightly different ($\ll$ 10\%) for the models with an extra heating, 
which can be explained by assuming that the DIG--layer is in pressure equilibrium. This effect was also noted by 
Wood \& Mathis~\cite{wood3}. The ionization parameters (U) of all models lie in the narrow range between 
log(U) = -3.0 and -3.1.

The ISM composition of CLOUDY is used with N/H and S/H set to the values
in Haffner99 (N/H = $7.5 \times 10^{-5}$, S/H = $1.86 \times 10^{-5}$). Graphite and silicate grains with the size distribution
used for the ISM (see HAZY, Ferland~\cite{ferland3}) are present in the gas and account for less than 10 \% of the global heating.
The inclusion of PAHs give only variations $<$ 3\% for the line ratios and have only a small effect on the heating balance.
The interaction with cosmic rays is taken into account as described in 
Ferland \& Mushotsky~\cite{ferland84}. \\

\begin{figure}
   \centering
   \includegraphics[angle=270,width=5.5cm]{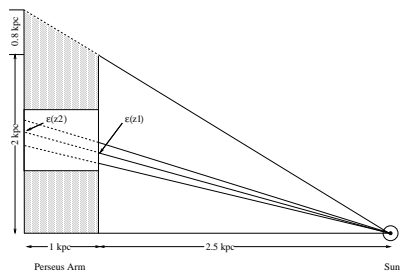}
      \caption{Sketch of the line of sight geometry and the beam smearing, which have to be considered
	when comparing the models with the data. Deriving temperatures from the data without taking this
	into account lead to temperatures that are on average 250 K higher.
              }
         \label{geometry-PerseusArm}
   \end{figure}

\begin{figure}
   \centering
   \includegraphics[width=5.5cm]{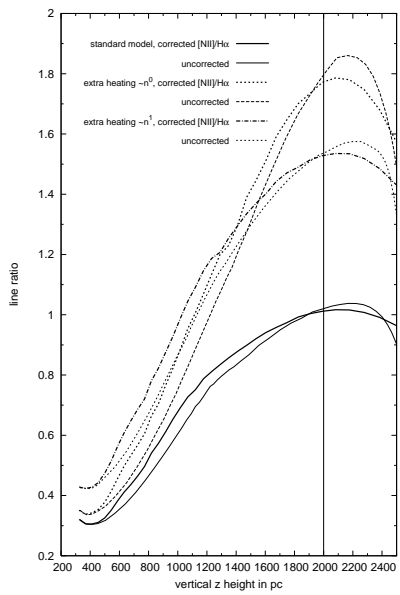}
      \caption{The effect of the beam smearing and line of sight geometry on [NII]/H$\alpha$ specifically.
	The enhancement of this line ratio due to this effect is at most 15\% and leads to higher temperatures 
	derived from the observations if the effects is not taken into account.
              }
         \label{geometry-beam_smearing}
   \end{figure}

A crucial factor is the consideration of the line of sight.
The observations give information about the line ratios at different positions
orthogonal to the source of the ionizing radiation. 
The correct local line ratios are represented by the ratios of the 
volume emission coefficients $\epsilon_V$: 
$ \frac{I_1}{I_2} = \frac{\int \; {\rm d}z \; \epsilon_{V_1}}{\int \; {\rm d}z \; \epsilon_{V_2}} \approx  \frac{\epsilon_{V_1}}{\epsilon_{V_2}}$.
An important issue is the treatment of the line of sight to the Perseus Arm
due to our position in the Milky Way as shown in Fig.~\ref{geometry-PerseusArm}.
Each pointing of the WHAM survey contains contributions from different
$|z|$ heights above the midplane ($\epsilon(z1)$ to $\epsilon(z2)$), this 
effect is taken into account by integrating over the particular sight line in the models.
The distance to the Perseus Arm is assumed to be 2.5 kpc and the thickness
of the arm to be 1 kpc as quoted in Haffner99.
Moreover, the observations are the average of a beam size of $1^\circ$ which
means that we have to take into account 
an additional integration over different z heights, on average 45 pc.
These effects of  geometry and 
observational smearing are taken into consideration here for the first time.
Fig~\ref{geometry-beam_smearing} shows the effect of the beam smearing and line of sight geometry on the calculated line ratios for z--heights
up to 2.5 kpc, which have to be considered for the observed heights of 1.8 kpc.
These effect increase the 'corrected' line ratios by at most 15\% and their trend is altered for high z--heights. 
The 'corrected' model without an extra heating seems to suggest that [NII]/H$\alpha$ is not increasing after about 
1.6 kpc, but the 'uncorrected' model shows a further increase due to the hardening of the radiation. 
The result of the gas temperature (section~\ref{comp}, Fig.~\ref{gastemp}) is even more influenced by this issue. 
The discussion of the line ratios therefore need to include an appropriate treatment of the line of sight geometry and the beam smearing to
take these effects into account.

\subsection{Extra heating}

Additional heating to photoionization is included which is proportional to
$n^1$ and $n^0$ in accordance with 
Reynolds et al. \cite{reynolds99}, their factors $G_1$ and $G_2$ respectively. We choose 
rates in the same range as in their paper: $G_1 = 1\times10^{-25}  \rm erg \, cm^{-3} \, s^{-1}$
and $G_2 = 5\times10^{-27}  \rm erg \, cm^{-3} \, s^{-1}$.
The heating--cooling balance can then be written as either $G_0 + G_1/n_e = \Lambda$
or $G_0 + G_2/n^2_e = \Lambda$. The heating due to photoionization is given by
$G_0 n^2_e$ and the cooling by $\Lambda n^2_e$.
The inclusion of an extra heating source rises the gas temperature of the models, at the same time
the ionization structure varies only slightly.
The temperature of the CLOUDY models are calculated by heating--cooling balance.
The extra heating therefore increases the temperature, having more pronounced effects at larger z heights as the photoionization heating
rate decreases like $\rm n^2$ as shown in Fig.~\ref{gastemp}. The graph is explained in
more detail in the next section.

In our models the ionization structure is nearly unaffected by the inclusion of an extra 
heating source, sulphur is slightly more effected than nitrogen. This behavior is
the basic assumption for the extra heating in Reynolds et al.~\cite{reynolds99}. However
they are assuming a constant ratio of $\rm N^+/N$, whereas the models show a dependence on
z height, this is expected as the radiation gets progressively absorbed.
Figure~\ref{ionfrac} shows ionization structure and the change  
dependent on the different heating rates. Hydrogen is nearly fully ionized throughout all models
which is a basic characteristic of the DIG.

\subsection{Comparison with Observations}\label{comp}

\begin{figure}
   \centering
   \includegraphics[width=5.5cm]{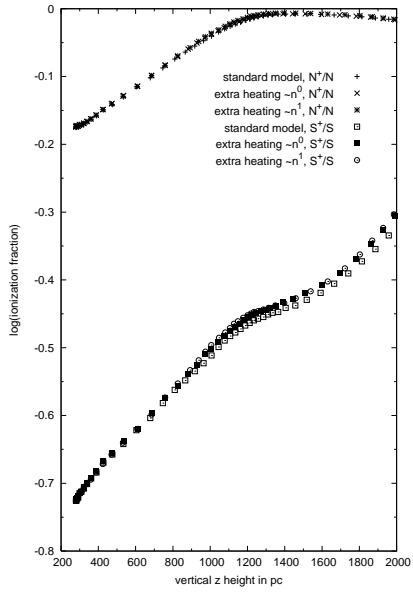}
      \caption{Ionization fractions of $\rm N^+$ and $\rm S^+$ dependent on 
        z, the assumption of Haffner99: $\rm N^+/N = H^+/H \rightarrow 1$ is approximately met for heights
	above 1300 pc.
              }
         \label{ionfrac}
   \end{figure}

\begin{figure}
   \centering
   \includegraphics[width=6cm]{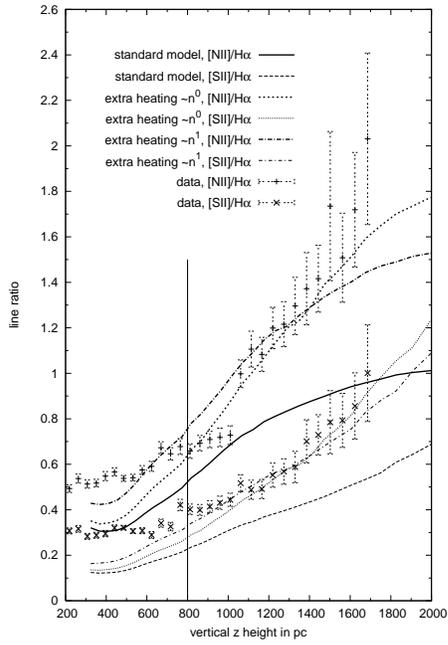}
      \caption{Observed [NII]/H$\alpha$ and [SII]/H$\alpha$ in the Perseus Arm, showing the
        characteristic rise. The models with and without an extra heating source are able to 
        explain this trend, the explicit values are better reproduced with an extra heating.
              }
         \label{n-h-z-PerseusArm}
   \end{figure}

\begin{figure}
   \centering
   \includegraphics[width=6cm]{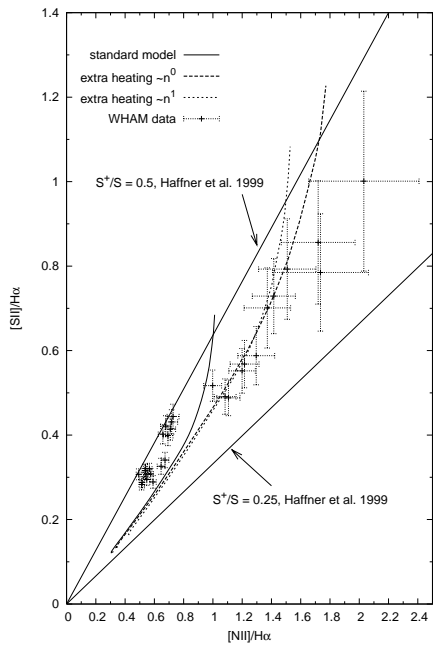}
      \caption{[NII]/H$\alpha$ -- [SII]/H$\alpha$ in the Perseus Arm as observed 
        by WHAM, plotted are the models, the data, and the limits given by Haffner99
	for two constant $\rm S^+/S$ ratios.
              }
         \label{sii-nii-PerseusArm}
   \end{figure}

In Fig.~\ref{n-h-z-PerseusArm} and Fig.~\ref{sii-nii-PerseusArm} the models differ by the type of extra
heating, ranging between models without extra heating and a rate of 
$G_1 = 1\times10^{-25}  \rm erg \, cm^{-3} \, s^{-1}$ and
$G_2 = 5\times10^{-27}  \rm erg \, cm^{-3} \, s^{-1}$.
There was no fitting done in order to match the models
with the observations. No 'best--fit' exists and therefore the
models are independent of the observational quality, individual spectral features, or
small scale variations which cannot be reproduced with a smooth density distribution.
As the data for small $|z|$ heights are contaminated with radiation from the midplane
and dust absorption it is convenient to consider line ratios for $|z|$--heights above 0.8 kpc to be 'pure' 
DIG. This is also the range for which the H$\alpha$ scale height was determined. The models show lower
values of [NII]/H$\alpha$ and [SII]/H$\alpha$ for the lower z heights as doubly ionized nitrogen contributes 35\% and
doubly ionized sulphur even 80\%, as seen in Fig.~\ref{ionfrac}. As a consequence [NII] and [SII] are weaker.
Figure~\ref{n-h-z-PerseusArm} shows the development of the line ratios with 
$|z|$ height, both the
data and the models show an increase with $|z|$. The trend in the line ratios is
matched by the models even without an extra heating source.

The [NII]/H$\alpha$ line
ratio above 1 kpc can be explained with the models including an extra heating rate. 
The modeled [SII]/H$\alpha$ ratio is up to a factor of two below the 
observations, however the theoretical uncertainty for the [SII] line
is very high as dielectronic recombination is an important process in the DIG.
As the corresponding recombination coefficients are not known 
(see discussion in Ferland et al.~\cite{ferland98}), the results of the
models have to be handled with care. Our models use the KLUDGE approximation (Ferland~\cite{ferland3}).
Models without dielectronic recombination have [SII]/H$\alpha$ decreased by 50\%. If the
rate is doubled then [SII]/H$\alpha$ is increased by 50\%.

\begin{figure}
   \centering
   \includegraphics[width=5.5cm]{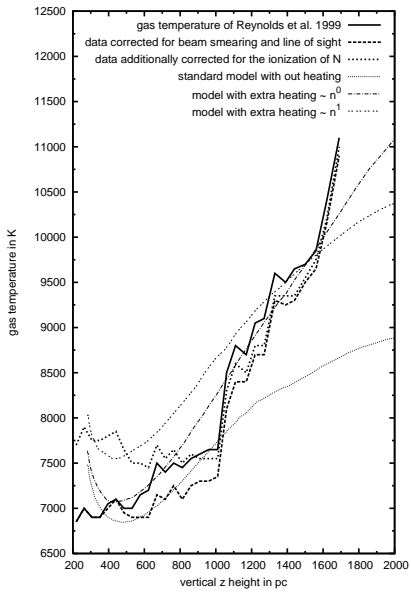}
      \caption{Temperature structure of the DIG, estimates from the observations with atomic data used by  
        Stafford et al.~\cite{stafford}. The line of sight geometry and beam smearing lead to temperatures
	on average 250 K lower, whereas the consideration of the ionization structure of $N^+$ elevates
	the derived temperatures again. The figure shows that a combination of the two different 
	heating terms gives the best agreement with the data.
              }
         \label{gastemp}
   \end{figure}

Figure~\ref{gastemp} shows the gas temperature of the models, the increase of the line ratios
with z height is due to the progressive hardening of the radiation as the photons go
through the gas layer. In order to explain the observed line ratios an extra heating source
is however needed which does not alter the general shape of the predicted line ratios 
but elevates the line ratios. As [NII] and [SII] are forbidden lines which get
collisional excited by electrons, an increase in gas temperature increases the amount of
electrons capable to excite the singly ionized nitrogen and sulphur ions which can then decay
by emitting the emission lines in questions.
The gas temperature is deducted from the observations through the relation - following Reynolds et al.~\cite{reynolds99}: 
$\frac{I_{[NII]}}{I_{\rm H\alpha}} = 1.84 \times 10^5 \left(\frac{\rm N^+}{N}\right) \left(\frac{\rm H^+}{H}\right)^{-1} T_4^{0.39} \exp(-2.18/T_4)$.
Assuming that $N^+/N = H^+/H$ and the abundance as stated in section~\ref{parameters} gives $\frac{I_{[NII]}}{I_{\rm H\alpha}} = 13.75 \, T_4^{0.39} \exp(-2.18/T_4)$, $T_4 = T / 10^4 \, \rm K$. 
We are not 
using the collision strength of singly ionized nitrogen from Reynolds et al.~\cite{reynolds99}: 
${\rm N^+}: \Omega\left(^3P,^1D\right) = 2.28\cdot T^{0.026}_4$, Aller~\cite{aller}, but the data from
Stafford et al.~\cite{stafford}: ${\rm N^+}: \Omega\left(^3P,^1D\right) = 3.02\cdot T^{-0.01}_4$. This
leads to the different coefficients and temperatures on average 250 K below the values of 
Reynolds et al.~\cite{reynolds99}. For singly ionized sulfur we use the data from
Lanzafame et al.~\cite{lanzafame93} instead of Aller~\cite{aller} as in Reynolds et al.~\cite{reynolds99}.
The plot also shows the impact of the line of sight geometry and 
beam smearing as well as the consideration of the ionization structure of the models. As the line of 
sight leads to higher [NII]/H$\alpha$ ratios (up to 15\%,) this effects, when accounted for, lower
temperatures. The ionization structure, i.e. $N^+/N \approx H^+/H$ is only valid for z $>$ 800 pc, leads to higher
estimates of the temperature. The combined effect is the elevation of the derived temperature for 
z $<$ 800 pc and above that lower values by about 200 K.
The models without an extra heating source have temperatures too low, an extra heating source independent of density shows
very good agreement with the observed temperatures. The extra heating $\propto n^1$ seems to best fit 
the data up to 800 pc, then the extra heating $\propto n^0$ gives the best agreement for higher
z values. The interpretation for the z heights $<$ 800 pc have to be handled with care as this region is 
contaminated by radiation from the midplane and therefore the part responsible for the DIG emission is
difficult to estimate.
Magnetic reconnection (e.g. Birk et al.~\cite{birk} ) or heating by cosmic rays through linear Landau--damping (Lerche \& Schlickeiser~\cite{lerche} )
are possible processes producing a heating independent of density. As photoelectric heating from
dust grains is $\propto n^1$, the data suggest that there is more dust present at z--heights below
800 pc than present in the models if this mechanism is responsible for the elevated temperature.

In Fig.~\ref{sii-nii-PerseusArm} the line ratios [NII]/H$\alpha$ and [SII]/H$\alpha$ are
plotted against each other. 
Values of [NII]/H$\alpha$ greater than 1, which cannot be explained by classical 
HII--region calculations, are reached with 
models using an extra heating source. Also in this case the extra heating independent of density is able
to produce higher [NII]/H$\alpha$ ratios than the other therefore matching better with the data.
The limits of Haffner99 for two constant $\rm S^+/S$ ratios (0.5 and 0.25) are also given. The ionization fractions of the models
(see Fig.~\ref{ionfrac}) are within these two limits for z $>$ 0.6$\,$kpc. Together with the temperature plot (Fig.~\ref{gastemp}) the
models match in all these direct and derived quantities with the data and agrees with the estimates of Haffner99 and 
Reynolds et al.~\cite{reynolds99}.
The application to other galaxies shows that this diagram is also a valuable diagnostic for the chemical evolution
(Elwert et al.~\cite{elwert}).

\section{Summary}\label{summary}

We have shown that the observed trend of the line ratios [NII]/H$\alpha$ and 
[SII]/H$\alpha$ above the Galactic plane can successfully be explained by 
photoionization models including extra heating and considering the line of sight geometry.
The observed values need an extra heating source which strength lies 
at the lower end of the predicted values of Reynolds et al.~\cite{reynolds99} due to our models. 
At high z heights (z $>$ 800 pc) an extra heating independent of density gives the best agreement with the data,
whereas for smaller z heights an extra heating term $\propto n^1$ gives better results concerning the 
temperature.
There is an intrinsically increase in the line rations and the gas temperature due to the 
progressive hardening of the radiation. The extra heating terms are enhancing this trend and
elevate the line ratios to the observed values.
It is important to incorporate the observing geometry to the Perseus Arm into the models
when comparing with the data.
A discussion concerning models of observed line ratios in edge--on galaxies is 
given in Elwert~\cite{elwert03} and Elwert \& Dettmar~\cite{elwert05}.

%________________________________________________________________
\acknowledgements
This work was supported by DFG through SFB 591 and
through Deutsches Zentrum f\"ur Luft- und Raumfahrt through grant 50 OR 9707.
TE wants to thank Kenneth Wood and Ron Reynolds
for helpful comments and enlightening discussions while writing the paper.
We also want to thank the anonymous referee for making many very useful suggestions
and comments which helped to improve the publication.
%________________________________________________________________

%________________________________________________________________

\clearpage

%________________________________________________________________
%------------------------------------------------------------------------------

\clearpage

%------------------------------------------------------------------------------
%------------------------------------------------------------------------------
\clearpage

%------------------------------------------------------------------------------
%------------------------------------------------------------------------------

\clearpage

%------------------------------------------------------------------------------
%------------------------------------------------------------------------------

\clearpage

%------------------------------------------------------------------------------
%------------------------------------------------------------------------------

\clearpage

%------------------------------------------------------------------------------
%------------------------------------------------------------------------------

\clearpage

%------------------------------------------------------------------------------
%------------------------------------------------------------------------------

\end{document}